\documentclass[sigconf,authorversion]{acmart}
\usepackage{hyperref}
\usepackage[utf8]{inputenc}
\usepackage{xcolor}
\usepackage{colortbl}
\hypersetup{hidelinks}
\usepackage{booktabs}
\usepackage{pifont}%
\usepackage{balance}

\newcommand{\cn}{\textsc{MultiMediate\,}}

\AtBeginDocument{%
  \providecommand\BibTeX{{%
    \normalfont B\kern-0.5em{\scshape i\kern-0.25em b}\kern-0.8em\TeX}}}

\copyrightyear{2024}
\acmYear{2024}
\setcopyright{acmlicensed}\acmConference[MM '24]{Proceedings of the 32nd ACM International Conference on Multimedia}{October 28-November 1, 2024}{Melbourne, VIC, Australia}
\acmBooktitle{Proceedings of the 32nd ACM International Conference on Multimedia (MM '24), October 28-November 1, 2024, Melbourne, VIC, Australia}
\acmDOI{10.1145/3664647.3689004}
\acmISBN{979-8-4007-0686-8/24/10}

\newcommand\philipp[1]{\textcolor{green}{}}
\newcommand\dominike[1]{\textcolor{orange}{}}
\newcommand\dominik[1]{\textcolor{red}{}}
\newcommand\michal[1]{\textcolor{blue}{}}

\begin{document}

\author{Philipp M\"uller}%
\affiliation{
    \institution{DFKI}
    \city{Saarbr\"ucken}
    \country{Germany}
}
\email{philipp.mueller@dfki.de}

\author{Michal Balazia}%
\affiliation{
    \institution{INRIA Sophia Antipolis}
    \city{Sophia Antipolis}
    \country{France}
}
\email{michal.balazia@inria.fr}

\author{Tobias Baur}%
\affiliation{
    \institution{University of Augsburg}
    \city{Augsburg}
    \country{Germany}
}
\email{tobias.baur@uni-a.de}

\author{Michael Dietz}%
\affiliation{
    \institution{University of Augsburg}
    \city{Augsburg}
    \country{Germany}
}
\email{michael.dietz@uni-a.de}

\author{Alexander Heimerl}%
\affiliation{
    \institution{University of Augsburg}
    \city{Augsburg}
    \country{Germany}
}
\email{alexander.heimerl@uni-a.de}

\author{Anna Penzkofer}%
\affiliation{
    \institution{University of Stuttgart}
    \city{Stuttgart}
    \country{Germany}
}
\email{anna.penzkofer@vis.uni-stuttgart.de}

\author{Dominik Schiller}%
\affiliation{
    \institution{University of Augsburg}
    \city{Augsburg}
    \country{Germany}
}
\email{dominik.schiller@uni-a.de}

\author{Fran\c{c}ois Br\'emond}
\affiliation{
    \institution{INRIA Sophia Antipolis}
    \city{Sophia Antipolis}
    \country{France}
}
\email{francois.bremond@inria.fr}

\author{Jan Alexandersson}
\affiliation{
    \institution{DFKI}
    \city{Saarbr\"ucken}
    \country{Germany}
}
\email{janal@dfki.de}

\author{Elisabeth Andr\'e}
\affiliation{
    \institution{University of Augsburg}
    \city{Augsburg}
    \country{Germany}
}
\email{elisabeth.andre@uni-a.de}

\author{Andreas Bulling}
\affiliation{
    \institution{University of Stuttgart}
    \city{Stuttgart}
    \country{Germany}
}
\email{andreas.bulling@vis.uni-stuttgart.de}

\title{\cn'24: Multi-Domain Engagement Estimation}
\date{July 1, 2024}

\renewcommand{\shortauthors}{Philipp Müller et al.}

\begin{abstract}

Estimating the momentary level of participant's engagement 
is an important prerequisite for assistive systems that support human interactions.
Previous work has addressed this task in within-domain evaluation scenarios, i.e. training and testing on the same dataset.
This is in contrast to real-life scenarios where domain shifts between training and testing data frequently occur.
With \cn'24, we present the first challenge addressing multi-domain engagement estimation.
As training data, we utilise the NOXI database of dyadic novice-expert interactions.
In addition to within-domain test data, we add two new test domains.
First, we introduce recordings following the NOXI protocol but covering languages that are not present in the NOXI training data.
Second, we collected novel engagement annotations on the MPIIGroupInteraction dataset which consists of group discussions between three to four people.
In this way, \cn'24 evaluates the ability of approaches to generalise across factors such as language and cultural background,
group size, task, and screen-mediated vs. face-to-face interaction.
This paper describes the \cn'24 challenge and presents baseline results.
In addition, we discuss selected challenge solutions.

\end{abstract}



\maketitle

\section{Introduction}

Knowing how engaged humans are in a conversation is an important prerequisite for many assistive systems, in particular if their aim is to maintain a high level of participation.
As a result, the estimation of human engagement has become an active research field, addressing a wide variety of approaches and scenarios.
These cover engagement prediction in human-human~\cite{muller2023multimediate,he2024tca}, and human-agent interactions~\cite{CharlesRichEngagementHumanRobotInteraction,park_model-free_2019, jain_modeling_2020}, as well as for different age groups including adults~\cite{muller2023multimediate,guhan_met_2022}, students~\cite{karimah_automatic_2021, goldberg_attentive_2021}, or children~\cite{rajagopalan_play_2015,oertel_engagement_2020}.
There is a wide variety of features that are utilised for engagement estimation, including backchannels~\cite{CharlesRichEngagementHumanRobotInteraction}, pose features~\cite{Sanghvi:2011:AAA:1957656.1957781BodyMotion}, or gaze data~\cite{Bednarik:2012:GCE:2401836.2401846Engagement}.
The inclusion of engagement estimation in the \cn'23 challenge lead to several new multi-modal engagement estimation approaches~\cite{yu2023sliding,he2024tca,yang2023multimediate,tu2023dctm}.

What all these approaches have in common is that they are trained and evaluated on the same dataset each.
While \citet{guhan_met_2022} applied their model trained on the MEDICA dataset on separately collected data, they did not evaluate engagement estimation performance on this separate dataset.
Despite clear progress on the engagement estimation task, within-domain testing does not reflect domain shifts that are frequent when applying approaches in the real world.
The complex nature of engagement makes it prone to influences of context variables.
Engagement might be expressed differently by people of different cultures, in different group compositions (dyadic vs. more than two people), and might be subject to different task characteristics.

In \cn'24 we pose the challenge of creating engagement estimation approaches that are able to transfer across such context factors.
To this end we significantly extend the engagement estimation task of \cn'23 by two new out-of-domain evaluation sets.
In particular, we introduce a not-yet released multilingual variant of the NOXI corpus~\cite{Cafaro:2017} to cover a wide variety of additional languages and cultural backgrounds, including Indonesian, Arabic, Spanish, and Italian.
Furthermore, we employ novel engagement annotations on the MPIIGroupInteraction corpus~\cite{muller_detecting_2018}, which consists of groups of three to four people engaged in face-to-face discussions.
Taken together, these evaluation sets vary along several dimensions: language and cultural background, group size, task, and screen-mediated vs. face-to-face interaction.
\cn'24 is embedded in a multi-year challenge with the goal of addressing several nonverbal behaviour analysis tasks relevant to autonomous artificial mediators. 
The first iteration of the challenge in 2021~\cite{muller2021multimediate} has addressed eye contact detection and next speaker prediction while \cn'22 focused on backchannel analysis~\cite{muller2022multimediate,amer2023backchannel}. \cn'23 addressed bodily behaviour recognition~\cite{muller2023multimediate,balazia2022bodily} and engagement estimation on the NOXI corpus~\cite{Cafaro:2017}.

In this paper, we define the multi-domain engagement estimation task, the evaluation criteria, and describe new annotations collected on the NOvice eXpert Interaction (NOXI) database~\cite{Cafaro:2017}, as well as on test and validation portions of MPIIGroupInteraction~\cite{muller_detecting_2018}.
Furthermore, we present baseline approaches for the challenge task and report evaluation results.
We make all collected annotations, baseline implementations, and raw feature representations publicly available for further use, also beyond the scope of \cn'24.\footnote{\url{https://multimediate-challenge.org}}

\section{Challenge Description}

In the following we present challenge task and the utilised training and testing datasets.
Testing data (without ground truth) was released to participants before the challenge deadline.
Participants in turn submitted their predictions for evaluation.

\subsection{Task definition}

In line with MultiMediate'23~\cite{muller2023multimediate}, the engagement estimation task addresses the frame-wise prediction of the conversational engagement level of each interlocutor on a continuous scale from 0 (lowest) to 1 (highest). 
To evaluate predictions on the test datasets, we make use of the Concordance Correlation Coefficient (CCC)~\cite{ccc} which ranges from -1 (perfect negative correlation) to 1 (perfect positive correlation).
The key difference to the engagement task in \cn'23 is that \cn'24 poses the challenge to address engagement estimation in a multi-domain evaluation scenario.
For this multi-domain evaluation, we employ two new out-of-domain test datasets: NOXI (Additional Languages) and the MPIIGroupInteraction test set.
Challenge participants are encouraged to develop methods that can generalise across these different domains and make use of multi-modal as well as reciprocal behaviour of both interlocutors.

\begin{figure}[t]
\centering
\includegraphics[width=1.0\columnwidth]{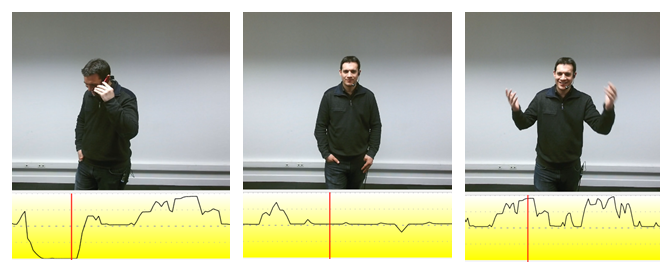}
\caption{A participant in the NOXI corpus being disengaged (left), neutral (center) and highly engaged (right).}
\label{fig:noxi:setup-5}
\end{figure}

\subsection{Datasets}

For the multi-domain engagement estimation task, we utilise three different datasets.
The training portion of the NOXI corpus~\cite{Cafaro:2017} is used as the main training dataset.
Testing is performed on three different test sets.
First, the in-domain NOXI test set which was already utilised in MultiMediate'23.
Second, an out-of-domain version of NOXI that includes conversations in languages not present in the NOXI training set.
Finally, we utilise test and validation portions of the MPIIGroupInteraction dataset~\cite{muller_detecting_2018} that were annotated with engagement labels specifically for this challenge.
In the following, we will present each dataset in detail.

\paragraph{NOXI}
For training, we follow \cn'23 and make use of the NOvice eXpert Interaction (\textsc{NOXI}) database \cite{Cafaro:2017,muller2023multimediate}.
NOXI is a corpus of dyadic, screen-mediated interactions in an expert-novice knowledge sharing context. 
In each session, one participant assumed the role of an expert and the other participant the role of a novice. 
Figure \ref{fig:noxi:setup-5} shows a user during interaction.
The goal of NOXI was to obtain data of spontaneous behaviour in a natural setting on a variety of discussion topics. Therefore, one of the main design goals was to match recorded participants based on their common interests. 
In a first step, potential experts were gathered
who expressed their willingness to share their knowledge about one or more topics they were knowledgeable and passionate about.
In a second step, novices were recruited that were willing to discuss or learn more about the available set of topics offered by experts.
The recording protocol furthermore introduced interruptions of the novices in order to provoke experts' reactions when conversational engagement gets interrupted.
NOXI includes interactions recorded at three locations (France, Germany and UK), spoken in eight languages (English, French, German, Spanish, Indonesian, Arabic, Dutch and Italian), 
discussing a wide range of topics. 
The dataset offers over 25 hours (x2) of interaction recordings, featuring synchronized audio, video (25fps), and motion capture data (using a Kinect 2.0). 
For training, and within-domain testing, we use a subset of this corpus containing 48 sessions for training and 16 sessions for testing (75/25 split). 
These sessions cover the languages English, French, German, and Dutch.
For \cn'23, each session was annotated in a continuous matter, meaning each video frame has a score between 0 and 1. Each rating was performed by at least two (up to 7) annotators (Average: 3.6 raters per session). We created gold standard annotations by calculating the mean over all raters. 
The NOXI dataset can be obtained from the website\footnote{\label{dataset-link}\url{https://multimediate-challenge.org/datasets/Dataset_NoXi/}}.

\paragraph{NOXI (Additional Languages)} 
This evaluation set includes four languages that are not part of the NOXI training set: two sessions in Arabic, two in Italian, four in Indonesian, and four in Spanish.
As a result, this evaluation set tests the ability of participants' approaches to transfer to new languages and cultural backgrounds not seen at training time.
We annotated these interactions for \cn'24, following the same protocal as in \cn'23~\cite{muller2023multimediate}.

\paragraph{MPIIGroupInteraction} 
To test the performance of participant's approaches in a different social situation, we make use of the MPIIGroupInteraction corpus\footnote{\label{dataset-link}\url{https://multimediate-challenge.org/datasets/Dataset_MPII/}}~\cite{muller_detecting_2018}.
This corpus consists of audiovisual recordings of group discussions between three to four participants, each lasting for 20 minutes.
MPIIGroupInteraction differs from NOXI in several key aspects.
First, it consists of group discussions while NOXI features dyadic interactions.
Second, there are no pre-defined roles such as ``novice'' or ``expert'', only the task to discuss on a topic that was selected to be controversial among the group members.
Third, interactions are face-to-face instead of screen-mediated.
Finally, participants are seated throughout the whole interaction.
With these differences to the NOXI setup, MPIIGroupInteraction presents a challenging evaluation case for engagement estimation approaches.
For \cn'23 we collected novel engagement annotations on the MPIIGroupInteraction test and validation sets.
The validation set with ground truth annotations is provided to participants to monitor their performance on the out-of-domain task.
In addition it may be used as a limited set of training data to develop supervised domain adaptation approaches.
The validation set comprises 6 recordings with 21 participants, while the test set consists of 6 recordings with 23 participants.

\begin{figure}[t]
  \centering
  \includegraphics[width=\columnwidth]{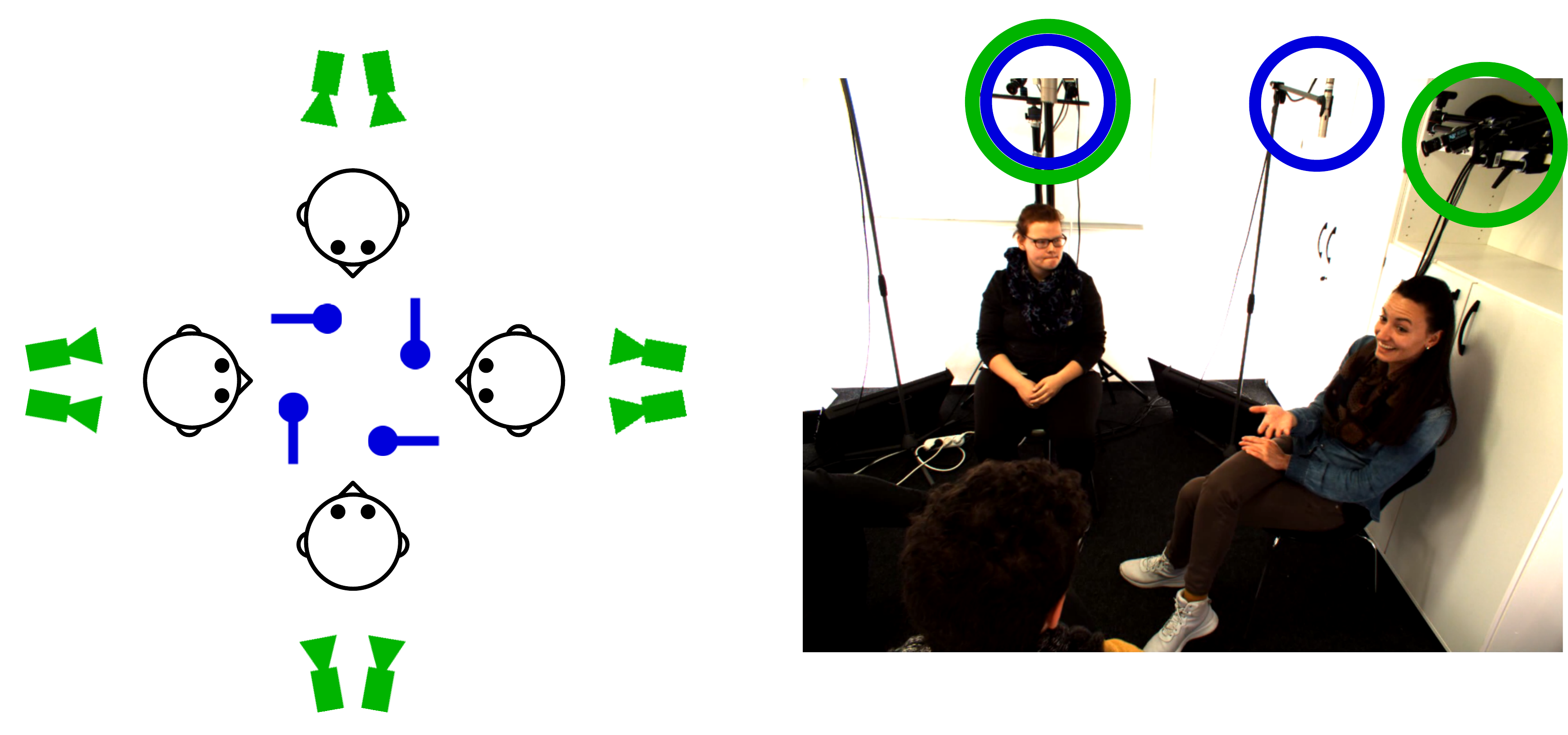}
  \caption{Setup of the MPIIGroupInteraction dataset. %
  Reproduced with permission from the authors of~\cite{muller_detecting_2018}.
  }~\label{fig:study_design_iui18}
\end{figure}

\section{Experiments}

We first present the different features we extracted on all datasets.
These features were used for our baseline experiments, and were also given to participants to develop their challenge solutions.

\subsection{Visual Features}

On both the NOXI and the MPIIGroupInteraction dataset, participants' locations or seating positions are known.
As a result, we can directly extract visual features, without the need to first localise and track participants.

\paragraph{Head Features} 
We extracted features from participants' head and face using OpenFace 2.0~\cite{baltrusaitis2018openface}. 
All features where extracted for each video frame.
The resulting feature vectors are consisting of 68 3D facial landmarks, 56 3D eye landmarks, presence and intensity of 18 action units as well as markers for detection success, detection certainty, facial position and rotation.

\paragraph{Pose Features} 
We extract body pose estimates using OpenPose~\cite{cao2017realtime}, resulting in a 139-dimensional feature representation including 2D body, hand, and facial keypoints for every video frame. %

\paragraph{CLIP Embeddings}
As a general visual representation, we extract CLIP (Contrastive Language-Image Pretraining)~\cite{radford2021learning} embeddings for each video frame. 
CLIP is trained to learn a joint embedding space for text and images.
In this way, it can capture a wide variety of semantic content present in the videos, e.g. relating to emotional expressions, attention, and many more.
The clip feature embeddings have 512 dimensions.

\begin{table*}[t]
\begin{tabular}{lcccccccc}
\toprule
& \multicolumn{2}{c}{NOXI} & \multicolumn{2}{c}{NOXI (Add. Languages)} & \multicolumn{2}{c}{MPIIGroupInteraction} & \multicolumn{2}{c}{Combined}\\
\cmidrule(lr){2-3} \cmidrule(lr){4-5} \cmidrule(lr){6-7} \cmidrule(lr){8-9}
Features             & Val CCC & Test CCC & Val CCC & Test CCC & Val CCC & Test CCC  & Test CCC\\
\midrule
\textit{Video}        &        &           \\
\ \ OpenFace 2.0      & 0.81   & 0.28     & -    & 0.13     & \textbf{0.09}    & 0.00  & 0.14\\
\ \ OpenPose          & 0.83   & 0.48     & -    & 0.41     & 0.01    & 0.06 & 0.32\\
\ \ CLIP              & \textbf{0.88}   & 0.48     & -    & 0.38     & -0.01    & 0.06 & 0.31\\
\midrule
\textit{Voice}       &       &          \\
\ \ eGemaps v2       & 0.77  & 0.56       & -    & 0.47     & 0.00    & \textbf{0.15} & 0.39\\
\ \ w2vbert2         & 0.77  & \textbf{0.64}       & -    & \textbf{0.51}     & 0.05    & 0.09 & \textbf{0.41}\\
\midrule
\textit{Text}       &       &          \\
\ \ XLM RoBERTa     & 0.62  & 0.40       & -    & 0.29     & 0.00    & 0.00 & 0.23\\
\bottomrule
\end{tabular}
\caption{Concordance correlation coefficient (CCC) of different featuresets on different engagement estimation validation and test sets. For NOXI (Additional Languages) no validation set is available.}
\label{tab:results_engagement}
\end{table*}

\subsection{Audio Features}
To analyse human speech for behavioural insights, it is essential to differentiate between its verbal and vocal components. 
The verbal aspect pertains to the use of words and language to express ideas, thoughts, or information. 
It encompasses the content of what is said. 
In contrast, the vocal component relates to the sounds produced by the voice, including tone, pitch, volume, and other characteristics of how something is said.
Both verbal and vocal features have been extracted using the DISCOVER framework \cite{schiller2024discover}.

\paragraph{Vocal Features} 
For the paralinguistic assessment of engagement, we extracted two sets of features using a one-second sliding window with a 40 ms stride, aligned with the video stream's frame rate. The first set is the Geneva Minimalistic Acoustic Parameter Set (eGeMAPS)~\cite{eyben2015geneva}, which includes 54 acoustic parameters frequently used in tasks such as depression, mood, and emotion recognition~\cite{valstar2016avec}. The second set of features was obtained using a pretrained w2v-BERT 2.0 encoder~\cite{barrault2023seamless}, which provides automatically learned representations of the audio signal. This model was trained unsupervised on a large dataset comprising 4.5 million hours of audio and has shown exceptional performance in various downstream tasks, including speech-to-text and expressive speech-to-speech translation.

\paragraph{Verbal Features}
To analyze the verbal content of spoken language, it is essential to first convert speech into text (STT). 
STT systems have been a focal point of research for many years. For our STT module, we utilize \textsc{WhisperX}~\cite{bain2022whisperx}, an adaptation of the \textsc{Whisper} Model \cite{radford2023robust}, which offers enhanced timestamp accuracy, support for longer audio sequences, and faster transcription.
When extracting features from text, the language of the text plays a crucial role. 
Since the challenge datasets are recorded in multiple languages, we employ multilingual textual feature extraction using the XLM RoBERTa model developed by \citet{unsupervised2019conneau}. 
This model is based on a transformer architecture and has been trained on a vast collection of multilingual data from the internet. 
To preserve the semantics of the transcript, every speech segment overlapping with the sliding window is included.

\subsection{Baseline Prediction Approach}

We evaluated the utility of the different feature modalities presented above for the task of frame-wise engagement estimation.
We implemented a fully connected neural network consists of an input layer followed by three hidden layers of size 136 each.
To prevent overfitting we relied on a dropout layer after the second hidden layer with a dropout rate of 0.25. 
The network was trained using the Adam optimizer and the mean squared error loss function. 
All hyperparameters were optimized using the hyperband search algorithm of the KerasTuner framework \cite{omalley2019kerastuner}.
We trained all approaches on the NOXI training set and evaluated on the three different test sets.
In particular, we did not use the MPIIGroupInteraction validation set for training.
Our baseline implementation is available online\footnote{\url{https://git.opendfki.de/philipp.mueller/multimediate24}}.

\section{Results}

We first discuss the results of our baseline experiments and then give an overview over the results achieved by teams participating in the challenge.

\subsection{Baseline Experiments}
The baseline results are depicted in Table \ref{tab:results_engagement}. 
The best average performance across all test sets is achieved by w2vbert2 features with an average CCC of 0.41, followed by eGemaps v2 features (0.39 average CCC).
Video features lack behind with OpenPose reaching the best performance at 0.32 average CCC.
Text features from XLM RoBERTa only achieve 0.23 average CCC.
All featuresets suffer from domain shifts.
This is especially severe when testing on the MPIIGroupInteraction dataset, where even the best approach (eGemaps v2) only achieved 0.15 CCC.
The impact of the domain shift on NOXI (Additional Languages) is less severe.
Nevertheless, all featuresets are impacted when comparing to the standard NOXI test set results.
For voice and text features, the degradation ranges from 0.09 CCC (eGempas v2) to 0.13 CCC (w2vbert2).
For visual features the impact has a similar range, from 0.07 CCC (OpenPose) to 0.15 CCC (OpenFace 2.0).
The fact that visual features are impaired indicates that in addition to speaking another language, there is also a shift in how visual nonverbal behaviour expresses engagement, pointing to the impact of different cultural norms that pose a challenge to generalisation.

\subsection{Challenge Solution Results}

\begin{table*}[t]
\begin{tabular}{clcccc}
    \toprule
    Rank & Approach & NOXI & NOXI (Add. Languages) & MPIIGroupInteraction & Combined \\
    \midrule 
    1 & USTC-IAT-United & 0.72 & \textbf{0.73} & \textbf{0.59} & \textbf{0.68}\\
    2 & AI-lab & 0.69 & 0.72 & 0.54 & 0.65 \\
    3 & \citet{li2024dat} & \textbf{0.76} & 0.67 & 0.49 & 0.64 \\
    4 & \citet{kumar2024} & 0.72 & 0.69 & 0.50 & 0.64 \\
    5 & ashk & 0.72 & 0.69 & 0.42 & 0.61 \\
    6 & YKK & 0.68 & 0.66 & 0.40 & 0.58 \\
    7 & Xpace & 0.70 & 0.70 & 0.34 & 0.58 \\
    8 & nox & 0.68 & 0.70 & 0.31 & 0.56 \\
    9 & SP-team & 0.68 & 0.65 & 0.34 & 0.56 \\
    10 & YLYJ & 0.60 & 0.52 & 0.30 & 0.47 \\
    \midrule
    11 & Baseline (ours) & 0.64 & 0.51  & 0.09 & 0.41\\
    \bottomrule
\end{tabular}
\caption{Challenge leaderboard for the engagement estimation task. Team names are replaced by references in case of accepted publications.}
\label{tab:leaderboard}
\end{table*}

We provide the challenge leaderboard for the multi-domain engagement estimation task in \autoref{tab:leaderboard}.
Approaches are ranked by the average test error across all test datasets.
In total, 10 approaches were able to surpass the baseline.
The best approach by the team USTC-IAT-United was able to reach an average CCC of 0.68, representing an increase over the baseline by 0.27 CCC.
Two papers describing challenge solutions were accepted for publication at ACM Multimedia~\cite{kumar2024,li2024dat}, both focusing on the multi-modal fusion mechanisms.
\citet{li2024dat} proposed to first fuse modality-specific representations across interactants and subsequently perform modality fusion.
This approach reached an average CCC of 0.64 and also set a new state-of-the-art on the in-domain NOXI test set with 0.76, outperforming TCA-NET which previously reached 0.75 CCC on the NOXI test set~\cite{he2024tca}.
\citet{kumar2024} in contrast proposed an approach which only processes features obtained from a single target participant.
They investigated different strategies to fuse modalities in a hierarchical fashion, reaching 0.64 average CCC in the challenge.
With respect to the cross-domain generalization abilities of participants' approaches, the gap between the original NOXI test set and NOXI (Additional Languages) is usually small, and in some cases non-existent (e.g. UST-IAT-United).
The approach of \citet{li2024dat} showed the largest gap, but also reached the highest performance on the original NOXI test set, indicating a larger degree of overfitting.
The domain gap between NOXI and MPIIGroupInteraction is still much larger.
Future approaches could investigate dedicated domain adaptation protocols to close this gap.

In addition to the multi-domain engagement estimation task, we also invited submissions to selected tasks from previous iterations of \cn.
An especially noteworthy method was proposed by \citet{ma2024less} who reached a new state of the art on the eye contact detection challenge~\cite{muller2018robust,muller2021multimediate}.
The authors proposed an adaptive feature selection method which can reduce computational burden while reaching high prediction accuracy.
They obtained an accuracy of 0.79, improving over the previous state of the art at 0.777 accuracy~\cite{li2023data}.

\section{Conclusion}
We introduced \cn'24, the first challenge addressing engagement estimation in a multi-domain evaluation scenario.
We presented novel annotations on publicly available datasets, including pre-computed feature representations. 
Furthermore, we defined the evaluation protocol, presented baseline results, and discussed successful challenge solutions.
Overall, we observed that while transfer between the original NOXI dataset and NOXI (Additional Languages) tends to work well, a larger domain gap remains when testing on MPIIGroupInteraction.
Datasets and evaluation will be accessible to researchers even beyond the \cn challenge, contributing to continuing progress on the challenge tasks.

\begin{acks}
P. M\"uller and J. Alexandersson were funded partially by the European Union Horizon Europe programme, grant number \grantnum{}{101078950}.
A. Bulling was funded by the European Research Council (ERC; grant agreement 801708). 
A. Penzkofer was funded by the Deutsche Forschungsgemeinschaft (DFG, German Research Foundation) under Germany’s Excellence Strategy – EXC 2075 – 39074001.
M. Balazia was funded by the \grantsponsor{}{French National Research Agency}{https://anr.fr/} under the UCA\textsuperscript{JEDI} Investments into the Future, project number \grantnum{}{ANR-15-IDEX-01}.
This work was supported in part by the Deutsche Forschungsgemeinschaft (DFG) through the Leibniz Award of E. André under Grant AN 559/10-1.
\end{acks}

\newpage
\bibliographystyle{ACM-Reference-Format}
\balance
\bibliography{bibliography}

\end{document}